\documentclass[11pt,twoside]{article}  
\usepackage{asp2006}
\usepackage{adassconf}
\usepackage{graphicx}

\begin{document}   

%
%

\paperID{O13.3}

%

\title{Transparent scientific usage as the key to success of the Virtual
Observatory}

%
%
%
%
%

\markboth{Chilingarian \& Zolotukhin}{}

%
%
%
%

\author{Igor Chilingarian\altaffilmark{1,2,3}}
\author{Ivan Zolotukhin\altaffilmark{3}}
\altaffiltext{1}{Observatoire de Strasbourg, CDS, CNRS UMR~7550,
Universit\'e de Strasbourg, 11 Rue de l'Universit\'e, 67000 Strasbourg, France}
\altaffiltext{2}{Observatoire de Paris, LERMA, CNRS UMR~8112, 61 Av. de
  l'Observatoire, 75014 Paris, France}
\altaffiltext{3}{Sternberg Astronomical Institute, Moscow State University, 13
Universitetskij prospect, 119992, Moscow, Russia}

%

\contact{Igor Chilingarian}
\email{igor.chilingarian@obspm.fr}

%
%
%

\paindex{Chilingarian, I.}
\aindex{Zolotukhin, I.}     

%



\begin{abstract}          
Nowadays, Virtual Observatory standards, resources, and services became
powerful enough to help astronomers making real science on everyday basis.
The key to the VO success is its entire transparency for a scientific user.
This allows an astronomer to combine ``online'' VO-enabled parts with
``offline'' research stages including dedicated data processing and
analysis, observations, numerical simulations; and helps to overpass one of
the major issues that most present-day VO studies do not go further than
data mining. Here we will present three VO-powered research projects combining VO and
non-VO blocks, all of them resulted in peer-reviewed publications.
\end{abstract}

%
%

\section{Introduction}

The Virtual Observatory is a realization of an e-Science concept in
astronomy. It forms a virtual environment aimed at facilitating
astronomical research and increasing scientific output of data by
providing transparent access to its resources including catalogues,
databases, archives, data visualization, processing and analysis tools.
There are conceptual similarities between the idea of the VO in astronomy,
and world-wide web in everyday life (Chilingarian 2009a).

The most important feature of the VO should be the transparency of this
infrastructure for a scientific user, who should be able to combine
VO and non-VO stages while working on his/her research task.

\section{Examples of VO-powered Research}

In this section we present three projects where we study galaxy properties
and evolution using Virtual Observatory technologies and tools. All these
examples resulted in publications submitted or published in major refereed
journals. For other examples of VO-powered research in Galactic astronomy
see, e.g. Zolotukhin (2010) and Zolotukhin et al. (2010).

\subsection{Search of Compact Elliptical Galaxies}

\begin{figure}
\includegraphics[width=0.3\hsize]{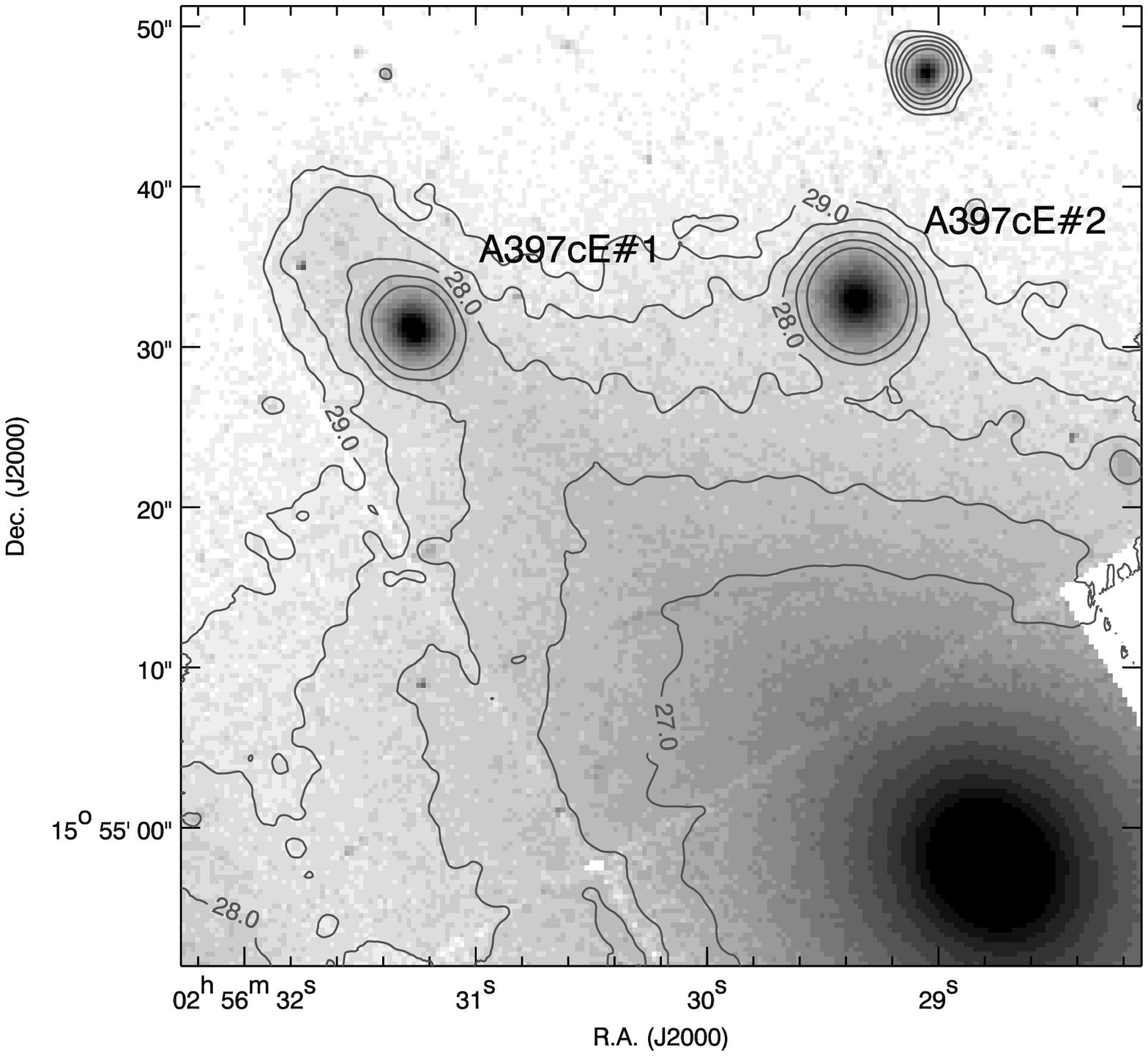}
\includegraphics[width=0.69\hsize]{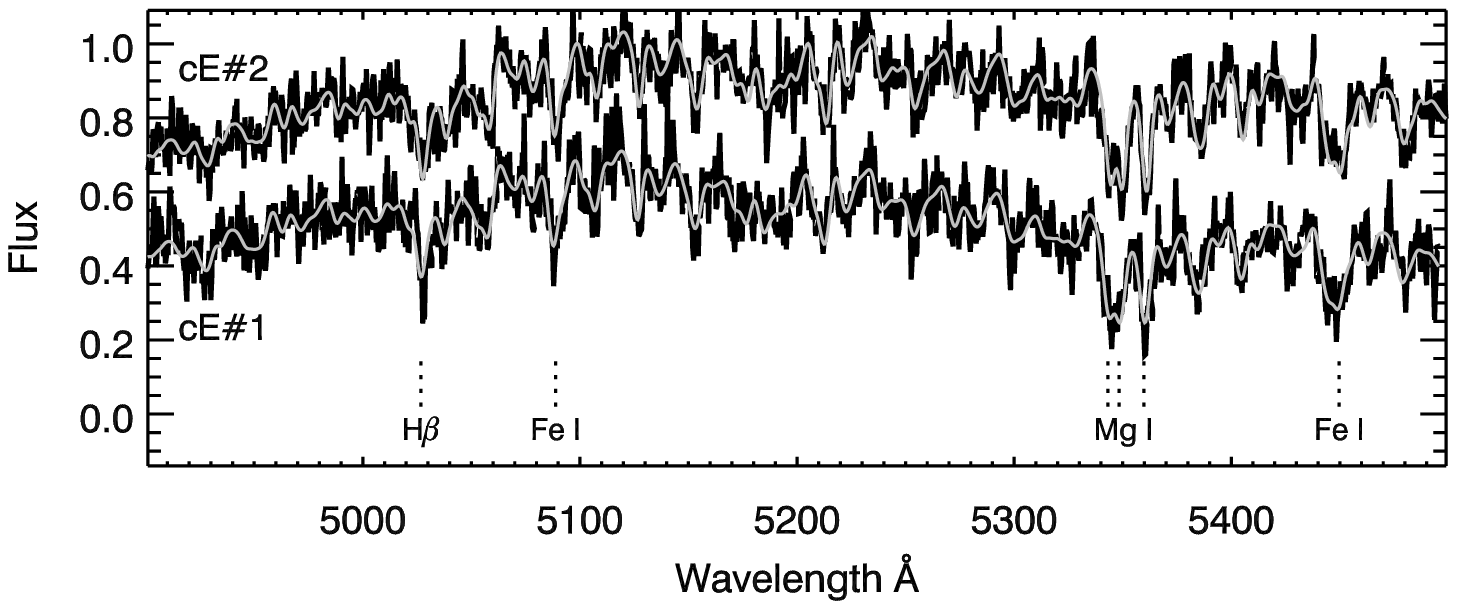}
\caption{A fragment of a WFPC2 HST image of
the central region of Abell~397 (left panel) with two candidate cE
galaxies. The A397cE\#1 candidate exhibits a
prominent extended low surface brightness tidal feature toward the
north-east. Optical spectra of
these two galaxies obtained with 6-m telescope are shown
in the right panel.\label{figcEA397}}
\end{figure}

Compact elliptical (cE) galaxies belong to a very rare class represented by
only a handful of confirmed members (Mieske et al. 2005, Chilingarian et al.
2007a, Price et al. 2009), such as the Andromeda galaxy satellite M~32. They
are characterised by small sizes (half-light radii $r_e \sim 0.1$~kpc) and
high stellar densities, and thought to form through tidal stripping of
massive progenitors, however, low statistics prevents us from making
decisive conclusions. The compact morphology makes them unresolved for
ground-based optical telescopes at a distances $>$50~Mpc. Using the superior
image quality provided by the Hubble Space Telescope (HST) would allow us to
push this limit to 200~Mpc, thus extending the volume of the Universe where
we can detect them by a factor of 60.

We designed and implemented a workflow to search for candidate cE galaxies
in large HST data collections provided by the Hubble Legacy
Archive\footnote{http://hla.stsci.edu/} and collect complementary
information using VO resources in order to confirm or reject them. We first
search for nearby clusters of galaxies ($z<0.055$) using the CDS Vizier
catalogue access service\footnote{http://vizier.u-strasbg.fr/} and NASA/IPAC
Extragalactic Database (NED)\footnote{http://nedwww.ipac.caltech.edu/}. Then
we use IVOA Simple Image Access Protocol to find and fetch HST images of the
galaxy clusters from the HLA. Each image is then processed using SExtractor
(Bertin \& Arnouts, 1996) in order to detect galaxies, obtain their
half-light radii and select compact objects with high surface brightness. At
the last step we query NED, Vizier and a spectral database of Sloan Digital
Sky Survey to find published redshifts or spectra of candidate objects.

Having applied the workflow to the HST WFPC2 data collection including
images of 63 nearby galaxy clusters, in 26 of them we detected 55 cE
candidates. For six of them we found archival SDSS DR7 spectra proving their
membership in the clusters, and for eight were confirmed with the redshifts
from literature. Spectra of seven other candidates in three galaxy clusters were
observed at the Russian 6-m telescope with the SCORPIO spectrograph
(Fig~\ref{figcEA397}). 

By applying the NBursts full spectral fitting (Chilingarian et al. 2007b) to
the available data for 13 galaxies, we determined velocity dispersions, ages
and metallicities of their stellar populations. All galaxies turned to have
high velocity dispersions and very old stars significantly more metal-rich
than what is observed in dwarf galaxies of similar luminosities (see e.g.
Chilingarian et al. 2008a, Chilingarian 2009b). All these properties
suggested the tidal stripping of intermediate-luminosity galaxies as a way
to create cEs. Similar situation is observed in ultra-compact dwarf galaxies
(Drinkwater et al. 2003, Chilingarian et al. 2008b) and transitional objects
(Chilingarian \& Mamon 2008).

In order to prove this hypothesis, we ran dedicated numerical simulations of
interactions of a disc galaxy with a galaxy cluster potential checking 32
different orbital configurations. Our simulations demonstrate the efficiency
of tidal stripping in reducing the stellar mass of a disc galaxy on a
timescale of 500--700~Myr by a factor of 2 to 10.

We converted the class of cE galaxies from ``unique'' into ``common under
certain environmental conditions''
The full description of the workflow and the astrophysical interpretation of
the discovery is available in Chilingarian et al. (2009a), the
first VO-based paper published in an interdisciplinary journal.

\subsection{FUV-to-NIR Properties of Low-Redshift Galaxies}

We have cross-identified three large sources of photometric data: GALEX
GR4 (UV), SDSS DR7 (optical), UKIDSS DR5 (NIR) and compiled a homogeneous
FUV-to-NIR catalogue of spectral energy distributions of nearby galaxies
($0.03<z<0.6$). We have extracted the data for the spectroscopically confirmed
galaxies and fitted their SDSS DR7 spectra to obtain stellar population
parameters, velocity dispersion and residual emission line fluxes of some
200000 galaxies. By using VO tools and technologies, all the computational
part of the study was completed in a week after the UKIDSS Data Release 5.
More details of this project are given in Zolotukhin (2010). The first paper
presenting the computation of $k$-correction, an essential step to construct
multi-wavelength SEDs, is submitted to MNRAS.

\subsection{The GalMer Database}

The GalMer Database\footnote{http://galmer.obspm.fr/} is a library including
thousands simulations of galaxy mergers at moderate resolution
(0.2--0.3~kpc), made available to users through tools compatible with the
Virtual Observatory (VO) standards adapted specially for this theoretical
database. To investigate the physics of galaxy formation through
hierarchical merging, it is necessary to simulate galaxy interactions
varying a large number of parameters: morphological types, mass ratios,
orbital configurations, etc. The GalMer database provides a reasonable
compromise between the diversity of initial conditions and the details of
underlying physics.

Apart from the direct access to simulations, we provide a set of value-added
services which allow users to compare the results of the simulations
directly to observations: stellar population modelling, dust extinction,
spectra, images, visualisation using dedicated VO tools. They can be used as
virtual telescope producing broadband images, 1D spectra, 3D spectral
datacubes, thus making this database oriented towards the usage by observers.

The paper presenting the GalMer database as a VO resource is submitted to
A\&A. However, the analysis of GalMer simulations and modelling their
stellar population properties have already been used to study the star
formation efficiency in interactions (Di Matteo et al. 2007, 2008b),
creation of old kinematically-decoupled systems (Di Matteo et al. 2008a),
reshaping metallicity gradients in early-type galaxies (Di Matteo et al.
2009), and some simulations were found to match a complex observed
light profile of a lenticular galaxy (Chilingarian et al. 2009b) proving its
origin from a major merger.

\section{Summary}

These examples aim at stimulating usual astronomers to carry out VO-enabled
research on everyday basis. We foresee a growing amount of VO-powered
studies to arrive in near future.

\acknowledgments
IC acknowledges the financial support by the VO-Paris Data Centre. IZ
thanks the ADASS POC for the allocated travel grant.

\end{document}